\documentclass[sor,reprint]{revtex4-1}
\usepackage{color,epsfig}
\usepackage{amsmath}
\usepackage{latexsym}
\usepackage{bm}
\usepackage{array,multirow}

\begin{document}

\title{New Global Potential Energy Surfaces of the ground $^3A'$ and $^3A''$ states of the H$_2$O system}

\author{Alexandre Zanchet}
\email[]{azanchet@ucm.es}
\affiliation{Departamento de Qu{\'\i}mica F{\'\i}sica I,
Facultad de Ciencias Qu{\'\i}micas, Universidad Complutense de Madrid, 28040 Madrid, Spain \\ Departamento de Qu{\'\i}mica F{\'\i}sica, Facultad de Ciencias
Qu{\'\i}micas, Universidad de Salamanca, 37008 Salamanca, Spain}

\author{Marta Men\'endez}\email[]{menendez@quim.ucm.es}
\affiliation{Departamento de Qu{\'\i}mica F{\'\i}sica I, Facultad de Ciencias
Qu{\'\i}micas, Universidad Complutense de Madrid, 28040 Madrid, Spain}

\author{Pablo G. Jambrina}
\email[]{pjambrina@usal.es}
%\thanks{}
\affiliation{Departamento de Qu{\'\i}mica F{\'\i}sica, Facultad de Ciencias
Qu{\'\i}micas, Universidad de Salamanca, 37008 Salamanca, Spain}

\author{F. Javier Aoiz} \email[]{aoiz@quim.ucm.es}
\affiliation{ Departamento de Qu{\'\i}mica F{\'\i}sica I, Facultad de Ciencias
Qu{\'\i}micas, Universidad Complutense de Madrid, 28040 Madrid, Spain}

\date{\today}

\begin{abstract}

This paper presents two new adiabatic, global potential energy surfaces (PESs) for the
two lowest $^3A'$ and  $^3A''$ electronic states of the O($^3P$)+H$_2$ system. For each
of these states, {\it ab initio} electronic energies were calculated for more than 5000
geometries using internally contracted multireference configuration interaction methods.
The calculated points were then fitted using the ansatz by Aguado {\it et al.} [Comput.
Phys. Commun. {\bf 108}, 259 (1998)] leading to very accurate analytical potentials  well
adapted to perform reaction dynamics studies. Overall, the topographies of both PESs are
in good agreement with the benchmark potentials of Rogers {\it et al.}, but those
presented in this work reproduce better the height and degeneracy of the two states at
the saddle point. Moreover, the long range potential in the entrance channel does not
require any cutoff. These features makes the new PESs particularly suitable for a
comparison of the dynamics on each of them. The new set of PESs were then used to perform
quantum mechanics and quasiclassical trajectory calculations to determine integral and
differential cross sections, which are compared to the experimental measurements by
Garton {\em et al.} [J. Chem. Phys., {\bf 118}, 1585 (2003)].
\end{abstract}


\maketitle %\maketitle must follow title, authors, abstract and \pacs

\section{Introduction}

The O($^3$P) + H$_2$ $\rightarrow$ OH($^2\Pi$) + H reaction has received a
great deal of attention over the last 60 years due to its importance in
combustion of hydrogen and hydrocarbons, where it plays a key role in chain
branching and propagation,\cite{TH:JCRD86,IR:C08} and in atmospheric reactions
in the upper atmosphere\cite{RD:GRL01}. It is also one of the most important
reactions in shocked interstellar clouds\cite{GD:AJ87} and in proto-planetary
disks.\cite{ACG:AA08} Consequently, there has been a long series of
measurements of thermal rate coefficients (see Ref.~\citenum{NS:JPCA14} and
references therein).

Although its detailed dynamics had been subject of numerous theoretical
studies,
\cite{B:CPL87,CFLTS:JCP93,Rogers:2000,HS:JCP00,HXW:CPL00,B:JCP04,WBBRW:JCP06,Z:JCP13,BWR:UJC15,ZL:CCC15}
experimental studies on the title reaction are scarce due to the large
electronic barrier of the reaction, requiring O($^3P$) atoms to be accelerated
to hyperthermal energies in a O($^1D$)-free atomic beam. Particularly relevant
for the reaction dynamics community are the experiments carried out by Minton
and coworkers,\cite{GMMTS:JCP03,GBMTMS:JPCA06,LZMM:NC13,LZMM:JACS14} where they
used a crossed-molecular beam machine coupled with a hyperthermal atomic-oxygen
beam source and laser-induced fluorescence detection to determine the
excitation function, the collision energy dependence of the integral cross
section (ICS),\cite{GMMTS:JCP03} differential cross sections (DCS),
\cite{GBMTMS:JPCA06} and rotational energy distributions
\cite{GBMTMS:JPCA06,LZMM:NC13,LZMM:JACS14} using laser induced fluorescence
with a resolution capable to determine the non-statistical $\Lambda$-doublet
population ratio of the nascent OH($^2\Pi$) as a function of the OH
vibrorotational state.\cite{LZMM:NC13,LZMM:JACS14}

Omitting the spin-orbit splitting, there are three potential energy surfaces
(PES) that correlate with the reactants in their ground states,  O($^3P_g$) +
H$_2(^1\Sigma^+_g)$; two of them, of symmetries $^3A'$ and $^3A''$, also
correlate adiabatically with the ground state products, OH($^2\Pi$)+H($^2S$).
These two PESs are degenerate for collinear geometries resulting from the $\Pi$
character of the linear arrangement of the three atoms. Given the fact that
\textit{ab-initio} calculations for both PESs systematically predict a
collinear saddle point with a barrier height close to 0.6 eV, one would expect
similar reactivities on the two PESs, at least at low collision energies.
However, as reactants move apart from a collinear geometry, the barrier height
rises more sharply in the case of the $^3A'$ state, which explains why $^3A''$
PES exhibits a larger reactivity. Of all the existing PESs available in the
literature, those calculated by Rogers \textit{et al.} \cite{Rogers:2000}
(hereinafter RWKW PESs) have been widely used in dynamical and kinetic
computational studies for the title reaction and they are supposed to be the
most accurate PESs to this moment. As such, they can be considered as benchmark
PESs for the title reaction.

In this article we primarily present the computational details of a new pair of global $^3A'$ and $^3A''$ PESs which have been recently
used to determine the $\Lambda$-doublet populations of the produced
OH(X).\cite{JZABA:NC16} High-level {\em ab initio} calculations have been
fitted to a convenient ansatz that includes analytical gradients, making them
very efficient to perform classical and semi-classical reaction dynamics
treatments. In fact, the collinear barrier, and most of the minimum energy
path, of the analytical PESs are virtually identical, in contrast to previous
sets of PESs.\cite{Rogers:2000}

This work is structured as follow.  Section II is devoted to the description of
the theoretical approaches,  the details on the {\it ab initio} calculations
are found in section \ref{abini}, the fitting procedure is described in section
\ref{fit} and the reaction dynamics calculations are presented in
section~\ref{Dyn}. The results are discussed in section~\ref{results} before
the conclusions in section~\ref{concl}.

\section{Theoretical methods}

\subsection{Ab initio calculations}
\label{abini}

{\it Ab initio} calculations were performed over the configuration space of the
OHH system to build  analytical PESs of the two lowest $^3A'$ and $^3A''$
states. All calculations were done using the MOLPRO suite of programs.
\cite{MOLPRO1,MOLPRO2} For both oxygen and hydrogen atoms, Dunning's
aug-cc-pV5Z basis set\cite{Dunning:89} including {\it spdfg} basis functions
were used. Both states correlate with both ground state asymptotic channels
O($^3P$)+H$_2$ and OH($^2\Pi$)+H($^2S$), where they are degenerate when
spin-orbit effects are not considered. To get an accurate and homogeneous
description of both PESs, the state-average complete active space (SA-CASSCF)
method\cite{Werner-etal:85} was employed to calculate the first $^3A'$ and the
two first $^3A''$ electronic states. The active space considered consists in 8
electrons distributed in 6 orbitals (2-6$\,a'$ and 1$a''$)  to include all
valence orbitals of oxygen and the 1{\it s} orbitals from both hydrogen atoms.
The resulting state-average orbitals and multireference configurations were
then used to calculate both the lowest $^3A'$ and the lowest  $^3A''$ states
energies with the internally contracted multireference configuration
interaction method (icMRCI) including simple and double
excitations\cite{Werner-etal:88} and Davidson correction.\cite{Davidson:75} The
$1s$ orbital of oxygen was kept frozen throughout the calculations.

To get a good description over the configuration space, the {\it ab initio}
points were sampled using the Jacobi coordinates ($r$, $R$, $\gamma$), where,
for a triatomic system A+BC,  ${\bm r=R}_{\rm BC}$, ${\bm R}$ is the vector
from the atom A to the center of mass of the diatomic BC, and $\gamma$ is the
angle between ${\bm r}$ and ${\bm R}$. Taking advantage of the symmetry of the
system, a first set of points have been calculated on a three-dimensional
regular grid ($N_r$=18) $\times$ ($N_R$=26) $\times$ ($N_\gamma$=10) on the
O+H$_2$ Jacobi coordinates, with (0.7 $\le r \le$ 4.9) $a_0$, (0.1 $\le R \le$
10) $a_0$ and (0 $\le \gamma  \le$ 90) degrees. To improve the description of
long range interactions in the O+H$_2$ channel, 20 values of $R$ (5 $\le  R
\le$ 10) $a_0$ have been added to the grid when (1 $\le r  \le$ 2) $a_0$.
Finally, a last set of points have been calculated on a three-dimensional grid
($N_r$=5) $\times$ ($N_R$=14) $\times$ ($N_\gamma$=7) based on the H+OH Jacobi
coordinates to refine the description of this asymptotic channel, with (1.6
$\le  r   \le$ 2.4) $a_0$, (2 $\le  R  \le$ 5) a.u. and (0 $\le \gamma  \le$
180) degrees. Overall, excluding non-converged points, there have been
calculated $\approx$5000 {\it ab initio} energies for each of the two PESs
sampled over all the configuration space. To complete the description, {\it ab
initio} calculations were also performed to describe the OH and H$_2$ diatomic
potential energy curves considering 40 and 52 points respectively.

\subsection{Fitting Procedure}
\label{fit}

The {\it ab initio} icMRCI+Q energies for the electronic states $^3A'$ and
$^3A''$, have been fitted separately using the GFIT3C procedure introduced in
Refs.~\cite{Aguado-etal:92,Aguado-etal:93,Aguado-etal:98}, in which the global
PES is represented by a many-body expansion:
\begin{equation}
V_{\rm ABC} = \sum_{\rm A} V_{\rm A}^{(1)} + \sum_{\rm AB} V_{\rm AB}^{(2)} (R_{\rm AB}) +
V_{\rm ABC}^{(3)} (R_{\rm AB},R_{\rm AC},R_{\rm BC}),
\end{equation}
where $V_{\rm A}^{(1)}$ represents the energy of the atoms (A=O,H,H)  in the ground
electronic state,  $V_{\rm AB}^{(2)}$ the diatomic terms (AB=OH,OH,HH)  and $V_{\rm
ABC}^{(3)}$ the 3-body term(ABC=OHH).

The diatomic terms can be represented by a sum of short- and long-range
contributions. The short-range potential is defined as a shielded Coulomb
potential,  whereas the long-range term is a linear combination of modified
Rydberg functions defined as:\cite{Rydberg:31}
\begin{equation}
 \rho_{\rm AB} (R_{\rm AB}) = R_{\rm AB} \, \exp\left[- \beta_{\rm AB}^{(2)}  \, R_{\rm AB} \right], \quad {\rm AB=OH,OH, HH}
\end{equation}
with $\beta_{\rm AB}^{(2)} > 0$. The root-mean-square (rms) error of the fitted
diatomic potentials  compared to the {\it ab initio} values are 4.0 and 1.7 meV
for OH and H$_2$, respectively. In a similar way, the 3-body term can also be
expressed as an expansion of modified Rydberg functions:
\begin{equation}
 V_{\rm ABC}^{(3)} (R_{\rm AB},R_{\rm AC},R_{\rm BC}) =
 \sum _{ijk}^K \, d_{ijk}\,  \rho_{\rm AB}^i \,\rho_{\rm AC}^j \,\rho_{\rm BC}^k
\end{equation}
This fitting procedure has the advantage of being fast in the evaluation of both the
potential and its analytical derivatives. Since symmetry of the system can be implicitly
included, GFIT3C is well adapted to fitting triatomic systems involving
H$_2$~\cite{Zanchet-etal:10,Dorta-Urra-etal:11,Zanchet-etal:13b,Dorta-Urra-etal:15}. This
property arises because for ABB systems there are only two non linear parameters to be
considered, $\beta_{\rm OH}$ and $\beta_{\rm HH}$ in the present case, which can be
coupled to additional constraints in the linear parameters $d_{ijk}$ to ensure symmetry
of the PES with respect to the permutation of the two H atoms, and thus reducing the
number of parameters to be evaluated.\cite{Aguado-etal:92,Aguado-etal:93,Aguado-etal:98}
The linear parameters $d_{ijk}$ (with $i+j+k \leq L)$  and the two nonlinear parameters
$\beta_{\rm OH}$ and $\beta_{\rm HH}$, are determined by fitting the calculated {\it ab
initio} energies after the substraction of the one- and two-body contributions. In the
present case, the order of the developmental $L$ was taken equal to 10 for both states.
One of the requirement of the three-body term is that it vanishes when one of the
internuclear distances reach infinity, but it can reach high values for short distances.
If the energy variation is too large, the fitting procedure tends to produce oscillations
due to the high degree of the polynomial chosen, leading to an unphysical description of
the long-range behavior which is required to be smooth. To avoid this problem, {\it ab
initio} points with high energies corresponding to geometries in the repulsive part of
the PES or near the total dissociation region have been excluded without significant loss
of precision as these regions are well described by the sum of diatomic potentials.
Finally, considering only 3400 and 3300 points (energies below 2.8 eV), an average rms
error of 11.7 meV (94 cm$^{-1}$) and 10.1 meV (81.5 cm$^{-1}$) have been obtained for the
$^3A'$ and $^3A''$ states, respectively.

\subsection{Dynamical Calculations}
\label{Dyn}

We have carried out time independent QM calculations on the PESs presented here using the
close-coupled hyperspherical method devised by Skouteris \textit{et al.} \cite{SCM:CPC00}
The results were obtained for a grid of 60 energies from $E_{\rm tot}$= 0.45 eV to
$E_{\rm tot}$= 2.5 eV for both parities of the incoming H$_2$ molecule and the triatomic
system. The basis set included all diatomic energy levels to $E_{\max}$= 3.25 eV. The
propagation was carried out in 300 log-derivative sectors until reaching a maximum
hyperradius of 15\,$a_0$ where the matching to product states is carried out. Total
angular momenta and up to 50 (62 at $E_{\rm tot} >1.5$\,eV). Helicities up to 15 and 24
have been included in the calculations on the $A'$ and on the $A''$ PESs, respectively.

Quasiclassical trajectories(QCT) were run using the procedure described in
Refs.~\citenum{ABH:JCSFT98,AHR:JCP92}. The excitation function was calculated by running
batches of 10$^6$ trajectories for each H$_2$ rovibrational state considered in this
work. For each batch, the collision energy was varied continuously between 0.35 and 2.0
eV. The trajectories were started at a atom-diatom distance of 10 \AA~ using an
integration step of 3$\times 10^{-17}$ s which guarantees a total energy conservation
better than one part in 10$^{5}$.

\section{Results and Discussion}
\label{results}

\subsection{Potential energy surface}
\label{PES}

\begin{figure}
\includegraphics[scale=0.35,angle=0]{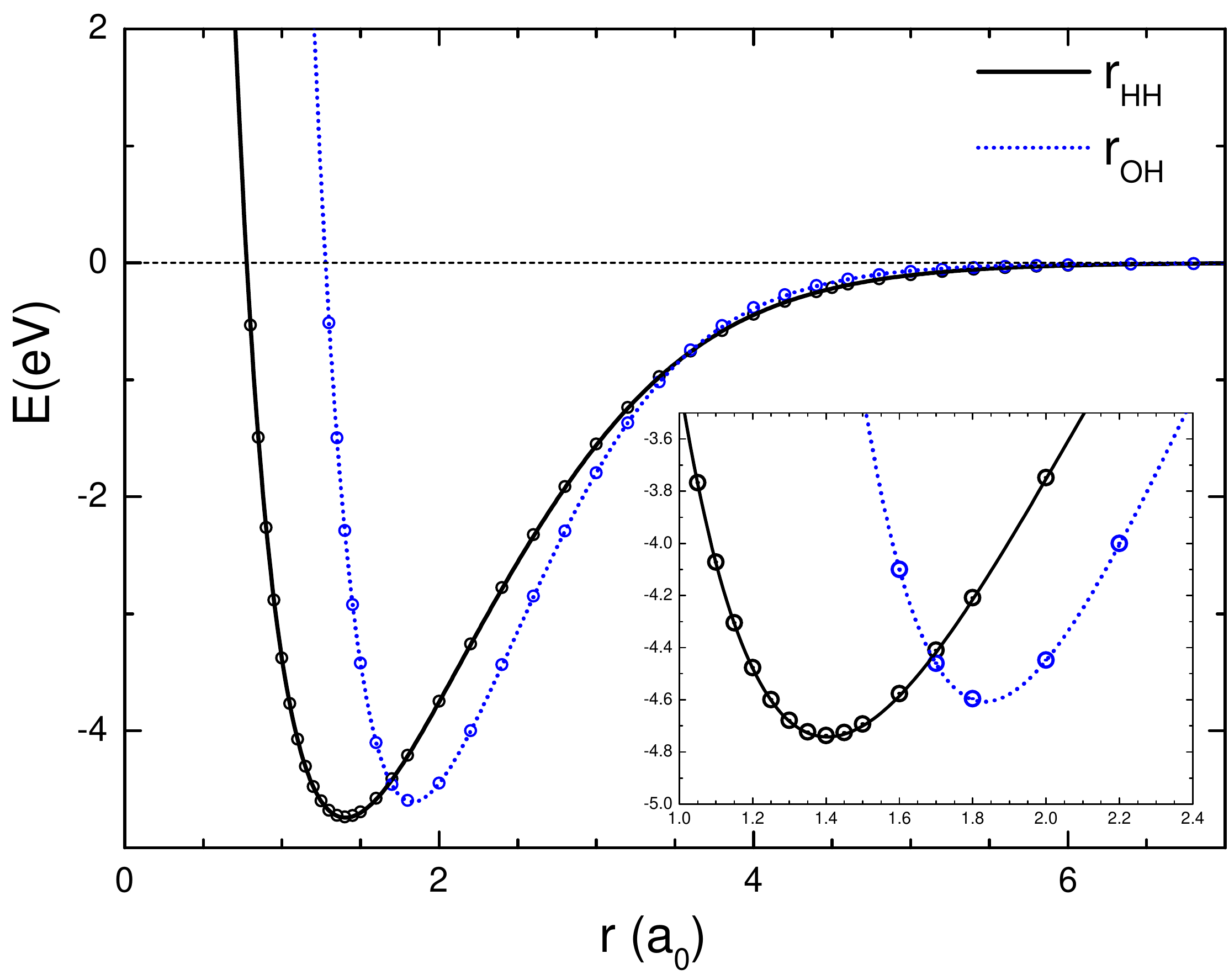}
\caption{Analytical potential energy curves for the
asymptotic H$_2$ (black line) and OH (blue line). Energies refer to the
respective dissociation in the ground state atoms. The circles along the curves
correspond to the {\em ab initio} results to assess the quality of the two-body
fits.}
\label{diatomic}
\end{figure}

As starting point, the diatomic potential curves for OH and H$_2$ are shown in
Fig.~\ref{diatomic}. The {\em ab initio} points are also included in the figure as open
circles to appraise the quality of the analytical fits corresponding to the two-body
terms. The main spectroscopic constants of each diatom in their ground states are shown
in Table~\ref{DiatConstant}, where they are compared with the experimental values. In
particular, the  experimental dissociation energies are very well reproduced by the
theoretical results. Using the diatomic potential curves, the
rovibrational energies have been calculated including all states below 3.25 eV total
energy. These data have been fitted to Dunham expansions, and the resulting spectroscopic
parameters are compared with the experimental data. The agreement is excellent for
$\tilde{\nu}_e$, $B_e$ and is fairly approximate for the anharmonicity constants,
considering that the experimental results are determined for a restricted set of
rovibrational states and that they do not include the $Y_{00}$ term.

%--------------------------------------------------------------------------
\begin{table}[ht]
\begin{center}
%\begin{ruledtabular}
\begin{tabular}{l|c|c|c|c|}
\hline \hline
& \multicolumn{2}{c|}{\quad\quad   H$_2$\quad\quad\quad}      &\multicolumn{2}{c|}{\quad\quad   OH\quad\quad\quad}      \\
\hline
                          &  Theor. &  Exp.    & Theor.   &  Exp. \\
\hline $D_e$  \quad             &  4.742  & 4.7483   & 4.607   &  4.5791 \\
$D^0_0$   \quad                 &  4.473  & 4.4781   & 4.376   &  4.392   \\
$r_e$     \quad                    &  1.403   & 1.4011  & 1.832   & 1.832   \\
$\tilde{\nu}_e$ \quad           & \quad 4400.71 \quad& \quad 4401.21 \quad  & \quad 3783.24 \quad & \quad 3737.76 \quad \\
$\tilde{\nu}_e \chi_e$ \quad    & 114.26  & 121.336  & 110.02  & 84.88  \\
$\tilde{\nu}_e y_e$   \quad     & -1.54   & 0.81     & 4.90    & 0.54    \\
$B_e$           \quad           & 60.33   & 60.87    & 18.89   & 18.91   \\
\hline \hline
\end{tabular}
\end{center}
\caption{Constants of the diatomic molecules. $D_e$ and
$D^0_0$ in eV; $r_e$ in $a_0$; $\tilde{\nu}_e$, $\tilde{\nu}_e \chi_e$,
$\tilde{\nu}_e y_e$ and $B_e$ in cm$^{-1}$. Theor., present work; Exp., from Ref.~\citenum{Huber1979}}
\label{DiatConstant}
%\end{ruledtabular}
\end{table}
%--------------------------------------------------------------------------

%------------------------------------------------------------
%------------------------------------------------------------------------------

Since we are interested in  O+H$_2$ collisions, the asymptotic energy of this channel is
taken as our zero reference energy. The main characteristics of the two triplet states
considered in this work are: (i) the absence of deep well able to stabilize the triatomic complex,
and (ii) the presence of two saddle-points (see Table~\ref{Saddle}). The lower
saddle-point corresponds to a linear O-H-H geometry where both states are degenerate.
{\it Ab initio} calculations  locate the saddle-point at $R_{\rm OH}$=2.301\,a$_0$ and  $R_{\rm
HH}$=1.681\,a$_0$ with an energy of 0.591 eV above entrance channel. It is thus the main
reaction barrier, located in the entrance channel of the O+H$_2$ collision, and identical
for both states. The fitted $A''$ and $A'$ analytical PESs reproduce the {\it ab initio}
saddle-point with very good accuracy and the degeneracy is accounted for almost exactly.
The saddle-point on the $A''$ analytical PES is found at $R_{\rm OH}=2.299$\,a$_0$ and
$R_{\rm HH}$= 1.681\,a$_0$ with an energy of +0.597 eV above the O+H$_2$ channel, while on
the $A'$ analytical PES, it is found at $R_{\rm OH}=2.295$ a$_0$ and $R_{\rm HH}$=
1.676\,a$_0$ with an energy of +0.597 eV. For comparison purposes, this saddle point  in
the RWKW PESs is at $R_{\rm OH}=2.300$\,a$_0$ and $R_{\rm HH}$= 1.706\,a$_0$ with an
energy of +0.565 eV for the $A''$ state and $R_{\rm OH}=2.310$ a.u. and $R_{\rm HH}$=
1.705\,a$_0$ with an energy of +0.573 eV for the $A'$ state. Hence, our global analytical
PESs overestimates the {\it ab initio} barrier height slightly, by 6 meV for both the
$A''$ and $A'$ states, while the RWKW PES underestimates the barrier by 26 and 18 meV,
respectively.\cite{Rogers:2000}

The second  saddle-point is found  at higher-energy for a linear H-O-H geometry, and is also common to both states. The {\it ab initio} saddle-point is found for both OH distances equal to
2.248\,a$_0$ and lies at +1.648 eV, as indicated in Table~\ref{Saddle}. Our analytical
fits reproduce the {\it ab initio} values well, as the saddle-point is found for $R_{\rm
OH}=2.250$\,a$_0$ and $R_{\rm OH}=2.239$\,a$_0$ for the $A''$ and $A'$ states, with 1.639
eV and +1.633 eV energies, respectively. For comparison with the RWKW PES, the
corresponding values are $R_{\rm OH}= 2.246$\,a$_0$ and $2.227$\,a$_0$, with energies of
+1.597 eV and +1.590 eV for $A''$ and $A'$ states, respectively. The present and RWKW
PESs are thus in good agreement for the description of the key regions of the potential,
although our PESs reproduce better the {\it ab initio} values and, moreover, the
degeneration of the two PESs for the collinear configurations.

%--------------------------------------------------------------------------
\begin{table}[ht]
\begin{center}
%\begin{ruledtabular}
\begin{tabular}{l|ccc}
\hline
\hline
                                    \multicolumn{4}{c}{OHH}    \\
\hline
& \quad \quad   $A'$\quad\quad       & \quad \quad   $A''$\quad\quad     & {\it ab initio}  \\
\hline
$R_{\rm HH}$ (a$_0$)                 &  1.676 & 1.681  &  1.682  \\
$R_{\rm OH}$ (a$_0$)                 &  2.295 & 2.299  &  2.301  \\
Energy (eV)                      &  0.597 & 0.597  &  0.591  \\
\hline
Stretching  (cm$^{-1})$ & 1756.6 & 1763.3 &         \\
Degenerate Bending  (cm$^{-1})$     & 789.3  & 460.8  &         \\
Stretching  (cm$^{-1})\quad $ & 1824.1$i$& 1810.9$i$&         \\
ZPE (eV)                         & 0.207  & 0.167  &         \\
Energy+ZPE (eV)                  & 0.804  & 0.764  &         \\
\hline
\hline
                             \multicolumn{4}{c}{HOH}    \\
\hline
$R_{\rm OH}$ (a$_0$)      &  2.239 & 2.240 &  2.248  \\
Energy (eV)               &  1.633 & 1.639 &  1.648  \\
\hline
\hline
\end{tabular}
\end{center}
\caption{\label{Saddle} Location and energies of the degenerated saddle-points for the
two analytical PESs compared with the optimized {\it ab initio} values. The energy is
given relative to the O+H$_2$ asymptotic channel. The harmonic frequencies of the
collinear saddle points and the resulting zero point energy (ZPE) are also given for both
analytical PESs. The bottom part of the Table correspond to the secondary H-O-H linear
saddle point.}
%\end{ruledtabular}
\end{table}
%--------------------------------------------------------------------------

%--------------------------------------------------------------------------
\begin{figure}
\includegraphics[scale=0.45,angle=0]{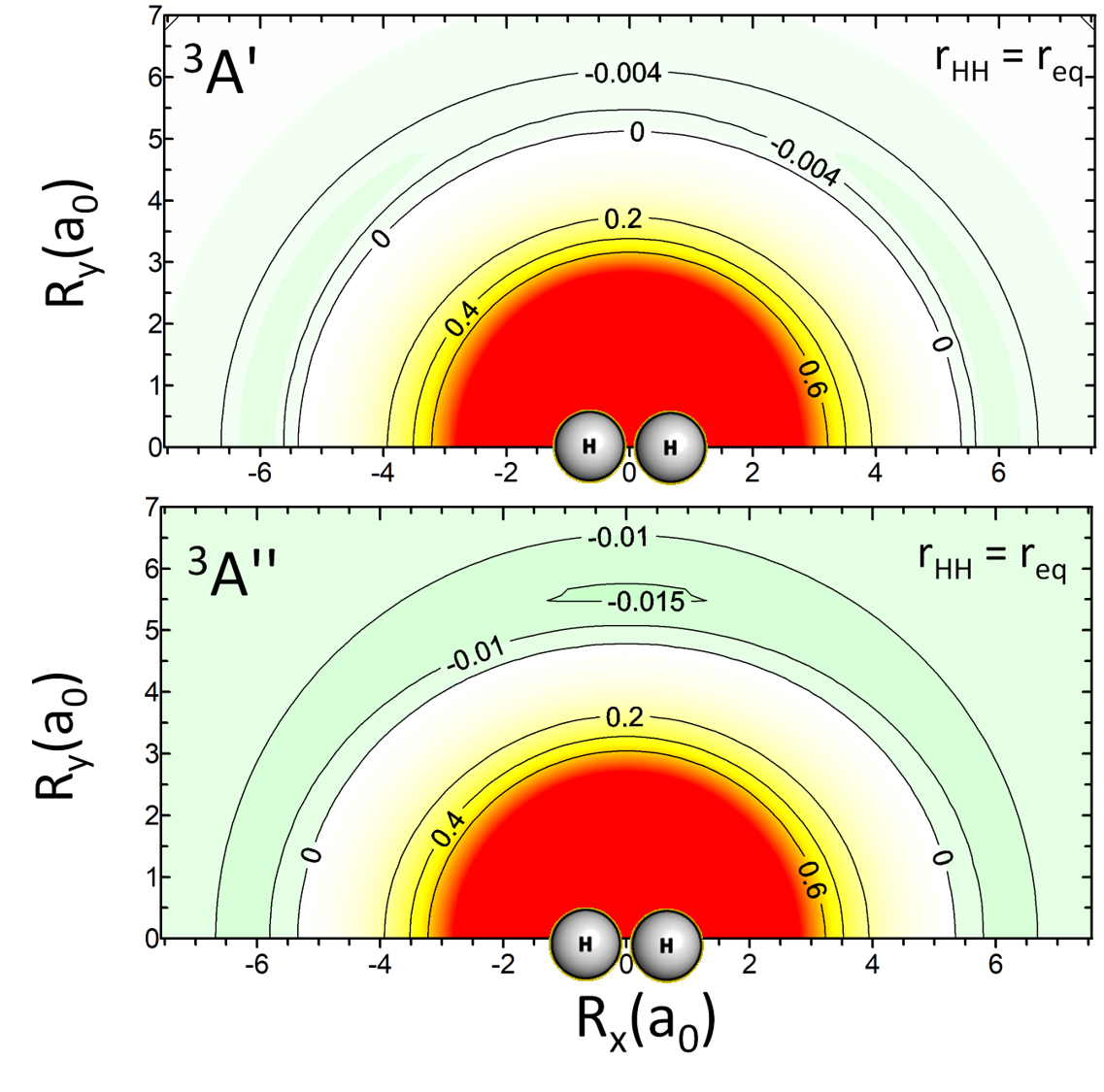}
% \vspace*{0.5cm}
\caption{Polar plot representation of the entrance
channel for H$_2$ at its equilibrium distance, $R_{\rm H-H}=$1.4\,a$_0$. The H$_2$ molecule
lies along the $x$-axis with its center in (0,0). Top panel corresponds to $A'$
state and bottom panel to $A''$ state.}
\label{polarentrance}
\end{figure}
%--------------------------------------------------------------------------

Although none of the $A'$ and $A''$ PESs feature deep wells able to form a stable
complex, both display  shallow van der Waals (vdW) wells in the O+H$_2$ entrance channel,
which are nearly isotropic with respect to the O--HH orientation.
For the $A''$ PES, the minimum of the vdW well is found for a T-shape geometry
while on the $A'$ PES, a linear configuration is preferred. Figure~\ref{polarentrance} shows the entrance channel potential topography of the vdW wells (in green) as polar $R_X$, $R_Y$ plot. As shown in Table~\ref{VDWwells}, the location of the wells and their depths are reasonably reproduced in the fitted PES when compared with the {\it ab initio} results. It is interesting to point out that the $A''$ features a slightly deeper well even at linear geometries.

\begin{figure}
\includegraphics[scale=0.35,angle=0]{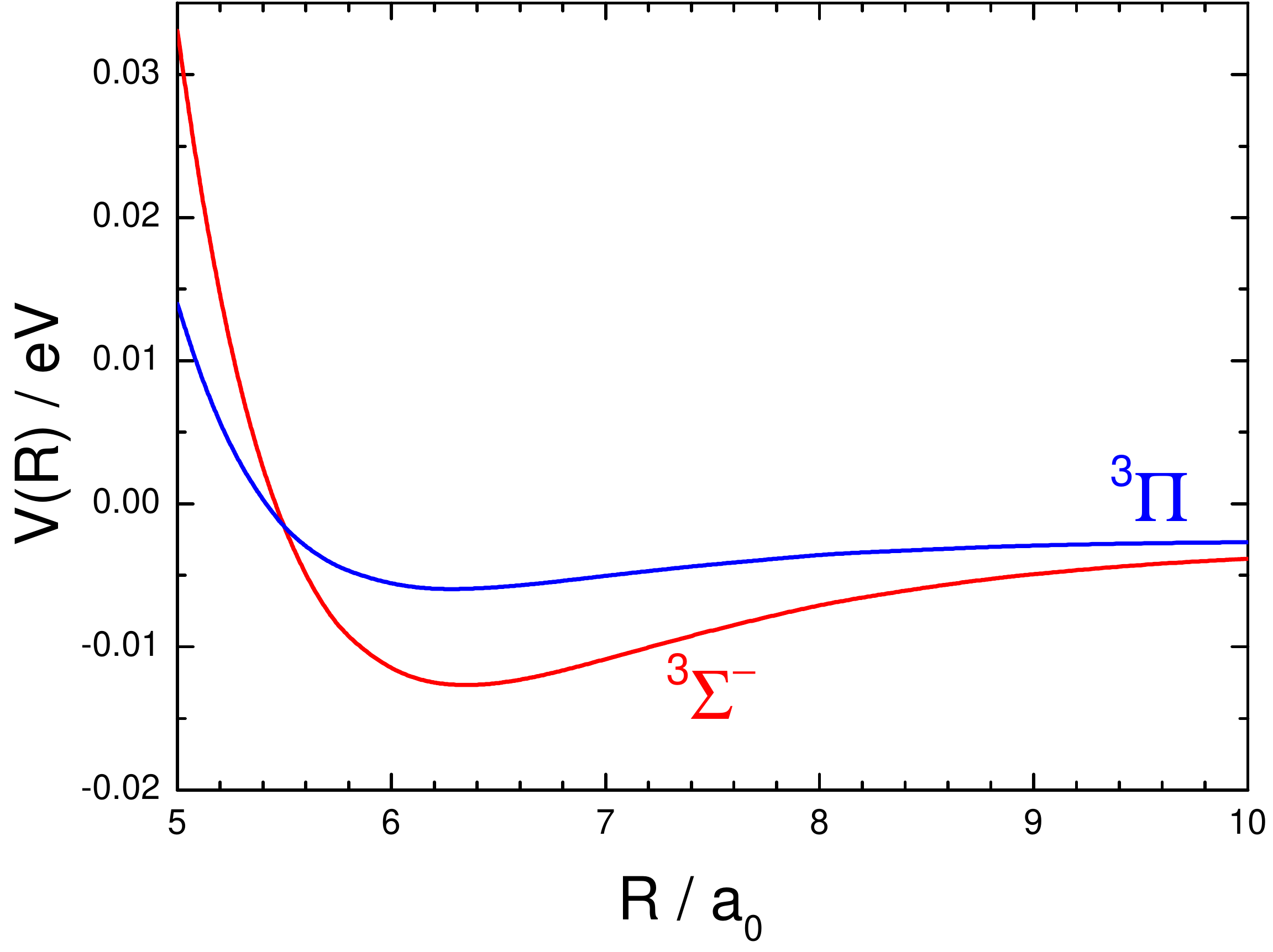}
% \vspace*{0.5cm}
\caption{Potential energy curve representing the $^3\Sigma^-$
and $^3\Pi$ diabats as a function of Jacobi vector {\bf R} in the
linear configuration for H$_2$ at its equilibrium distance, $R_{\rm HH}$=1.4\,a$_0$}
\label{crossing}
\end{figure}

This difference in linear configuration between the two states is due to fact that in the
long range region where the wells appear, the $^3\Sigma^-$ state is lower in energy than
the $^3\Pi$ state, as can be seen in Fig.~\ref{crossing}. For $R_{\rm H_2}$ at its
equilibrium distance (1.4\,a$_0$), both states cross at $R\approx$5.5\,a$_0$, leading
to a conical intersection associated to Renner-Teller couplings. Since in this work we
are interested in the construction of analytical PESs of the adiabatic states,
Renner-Teller couplings were not studied in detail and the conical intersection is
smoothed by the fitting procedure so that the resulting PESs were exempt of
singularities. For larger distances, $^3\Sigma^-$ state (correlating to $A''$ in $C_s$
symmetry) is more stable and  thus the $A''$ state is lower in energy than $A'$.
In contrast, for shorter distances (near the saddle-point), the $^3\Pi$ state is lower in
energy and both $A'$ and $A''$ states become strictly degenerate in linear configuration.
We should remark here that this long range region, where $^3\Sigma^-$ is the ground
state, was not included in the RWKW PES, and the description of this region is one of the
major difference between the present set of PESs and those from
Ref.~\citenum{Rogers:2000}. This difference in the entrance channel of both states may
have significant effect for inelastic collisions at low energies.\\

In the OH channel, both states  present two degenerate VdW wells in linear
configuration. The H--OH well lies at an energy of 117 meV above the O+H$_2$
channel in the {\it ab initio} calculation (depth of -18 meV with respect to
H+OH asymptote) and corresponds  to a linear configuration for OH distances of
1.833\,a$_0$ and 6.011\,a$_0$ (see Table~\ref{VDWwells}). The fit of the $A''$
PES reproduces this well, but overestimating its lying energy by 27 meV, while
the fit of the  $A'$ PES  exhibits no minimum for this configuration. This is
the only feature that it is not well described by the new set of PESs. For the
H--HO well, both fits reproduce accurately the vdW wells  arising for OH
distance of 1.835\,a$_0$ and HH distance of 4.299\,a$_0$ lying at -14 meV below
the OH+H  channel.

%----------------------------------------------------------------------------
\begin{table}[bht]
\begin{center}
%\begin{ruledtabular}
\begin{tabular}{l|cc|cc}
\hline \hline
                         \multicolumn{5}{c}{\rm O-HH}                          \\
\hline
                     &     \multicolumn{2}{c}{$A'$}   & \multicolumn{2}{c}{$A''$}  \\
\hline
                     & \quad\quad   Fit\quad\quad\quad     & {\it ab initio}   & \quad\quad   Fit\quad\quad\quad     & {\it ab initio}  \\
\hline
$r$(a$_0$)            &    1.401    &       1.402          &   1.401  &   1.403        \\
$R$ (a$_0$)          &  5.741   &     5.762         &  5.620 &      5.553       \\
$\gamma$ (deg.)      &   90     &       90          &   90   &      90          \\
Energy (eV)          & -0.005   &     -0.009        & -0.015 &     -0.013       \\
\hline \hline
                         \multicolumn{5}{c}{O-HH linear}                          \\
\hline
%                     &     \multicolumn{2}{c}{$A'$}   & \multicolumn{2}{c}{$A''$}  \\
%\hline
%                     &  Fit   & {\it ab initio}   &  Fit   & {\it ab initio}  \\
%\hline
$R_{\rm HH}$ (a$_0$)       & 1.402  &    1.401    &  1.403  &  1.402      \\
$R_{\rm OH}$ (a$_0$)       & 6.002  &     5.592   &  6.160  &  5.630      \\
Energy (eV)                & -0.006 &    -0.006   & -0.012  &  -0.013     \\
\hline \hline

                                   \multicolumn{5}{c}{\rm H-OH}                  \\

\hline
$R_{\rm HH}$ (a$_0$)       &  -  &  7.844         & 7.755          &   7.844          \\
$R_{\rm OH}$ (a$_0$)       &  -  &  1.833         & 1.832          &   1.833          \\
$\widehat{\rm HOH}$ ($^\circ$)     &  -  & 180            &  180           &     180          \\
Energy (eV)         &   -  & 0.117    & 0.131    &  0.117   \\
                    &   -  & (-0.018) & (-0.004) &  (-0.018)  \\

\hline \hline
                                   \multicolumn{5}{c}{\rm H-HO}                  \\
%\hline
%                     &     \multicolumn{2}{c}{$A'$}   & \multicolumn{2}{c}{$A''$}  \\
%\hline
%                     &  Fit   & {\it ab initio}   &  Fit   & {\it ab initio}  \\
\hline
$R_{\rm HH}$ (a$_0$)       &  4.340 &  4.299         & 4.305          &   4.299          \\
$R_{\rm OH}$ (a$_0$)       &  1.834 &  1.835         & 1.834          &   1.835          \\
$\widehat{\rm HHO}$ ($^\circ$)  &  180   & 180            &  180           &     180          \\
Energy (eV)          &   0.121(6) & 0.121  & 0.121(4)  &  0.121   \\
                     & (-0.013) &(-0.014)& (-0.013) &  (-0.014)  \\

\hline \hline
\end{tabular}
\end{center}
\caption{\label{VDWwells} Characteristics of the Van der Waals wells of two analytical
PESs compared with the optimized {\em ab initio} values. Energies are given relative to
the O+H$_2$ asymptotic channel, and for completeness, energies relative to the OH+H
asymptote are given in parenthesis for the H-OH Van der Waals well.}
%\end{ruledtabular}
\end{table}
%-------------------------------------------------------------------------------------------------

%-------------------------------------------------------------------------------------------------
%-------------------------------------------------------------------------------------------------
\begin{figure}
\hspace{0.5cm}
\includegraphics[scale=0.38,angle=0]{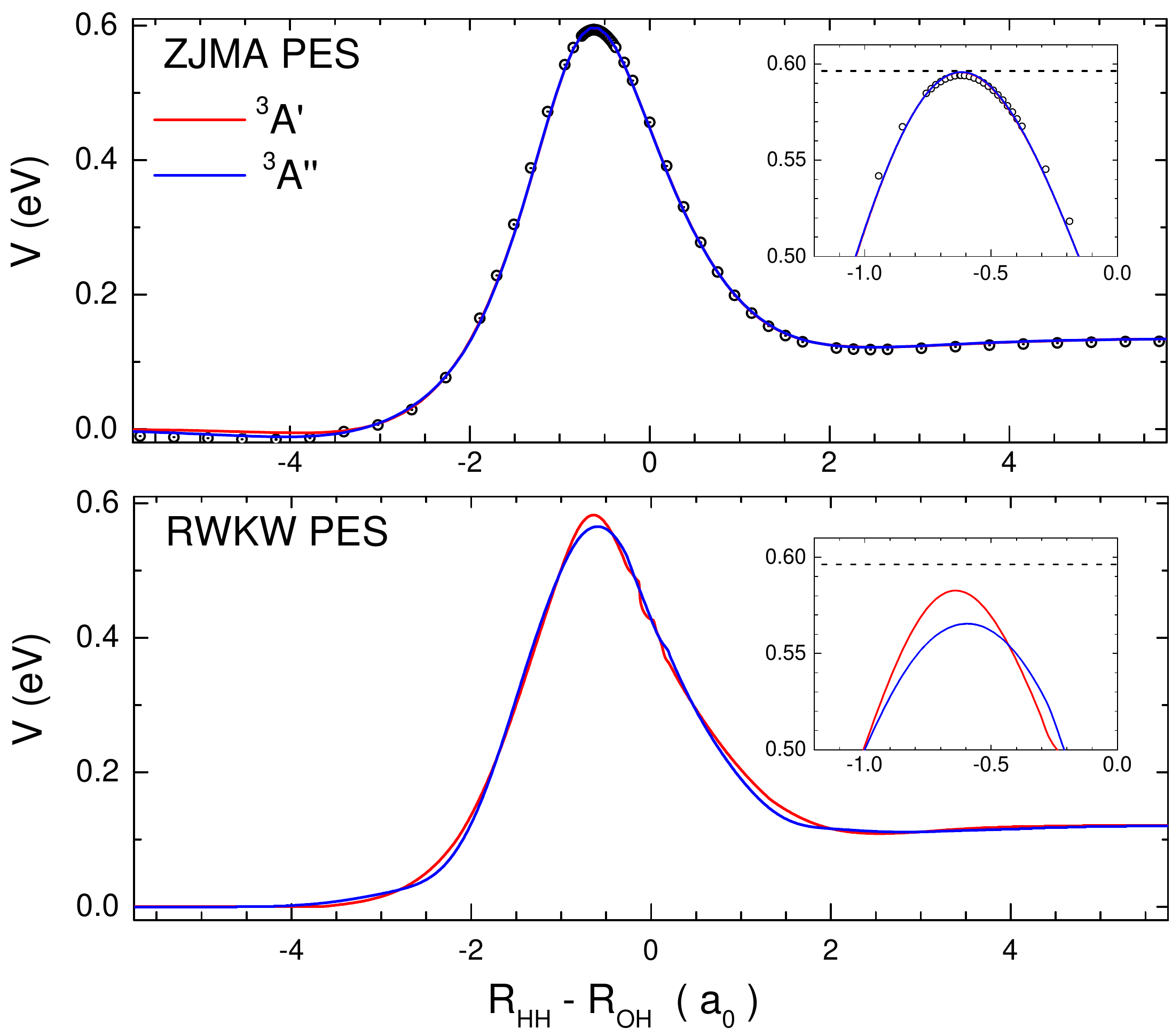}
% \vspace*{0.5cm}
\caption{Minimum energy path in collinear geometry as a
function of $R_{\rm HH}$-$R_{\rm OH}$. The top panel corresponds to the current
PESs and the bottom panel to the RWKW PESs. As highlighted in the inset,
the barrier heights of both PESs of the present work are almost identical and in excellent agreement with the (degenerate) {\em ab initio} points.At the entrance channel, $R_{\rm HH}-R_{\rm OH} < -4$\,a$_0$, where the two states are not degenerate (see Fig.~\ref{crossing}), the represented points correspond to the $^3A''$ PES. On the fitted RWKW PESs, in contrast, the differences between the saddle points are evident. To ease the comparison between the two sets of PESs, a horizontal line corresponding to the barrier height for the current $^3A'$ PES is shown in
both panels.}
\label{meplinear}
\end{figure}

If the system is kept at a linear configuration, where the reaction barrier is lowest, the Minimum Energy Paths (MEPs) of the $A'$ and $A''$ fits, shown in
Fig.~\ref{meplinear}, are almost identical since both states are degenerate in the region of the saddle point. There is, however a perceptible difference in the entrance channel due to the crossing between the $\Sigma^-$ and $\Pi$ states mentioned previously. The MEPs calculated on the present PESs are very similar to those obtained using the RWKW PES (bottom panel of Fig.~\ref{meplinear}), albeit with some appreciable differences. As shown in the insets of Fig.~\ref{meplinear}, where the respective transition state regions are highlighted, the barrier is slightly higher on the present PESs, and it is in an almost perfect agreement with the {\em ab initio} values. In addition, the $A'$ and the $A''$ degeneracy in the whole region of the collinear transition state is extremely well preserved in the present analytical PESs. In contrast, the barrier on the RWKW $A'$ lies higher than on the
$A''$ and the location of the transition state is slightly shifted towards the entrance channel on the former. Clearly, the present PESs are thus more precise to study the comparative reactivity on the $A'$ and $A''$ PESs.

%-----------------------------------------------------------------------
\begin{figure}
\includegraphics[scale=0.35,angle=0]{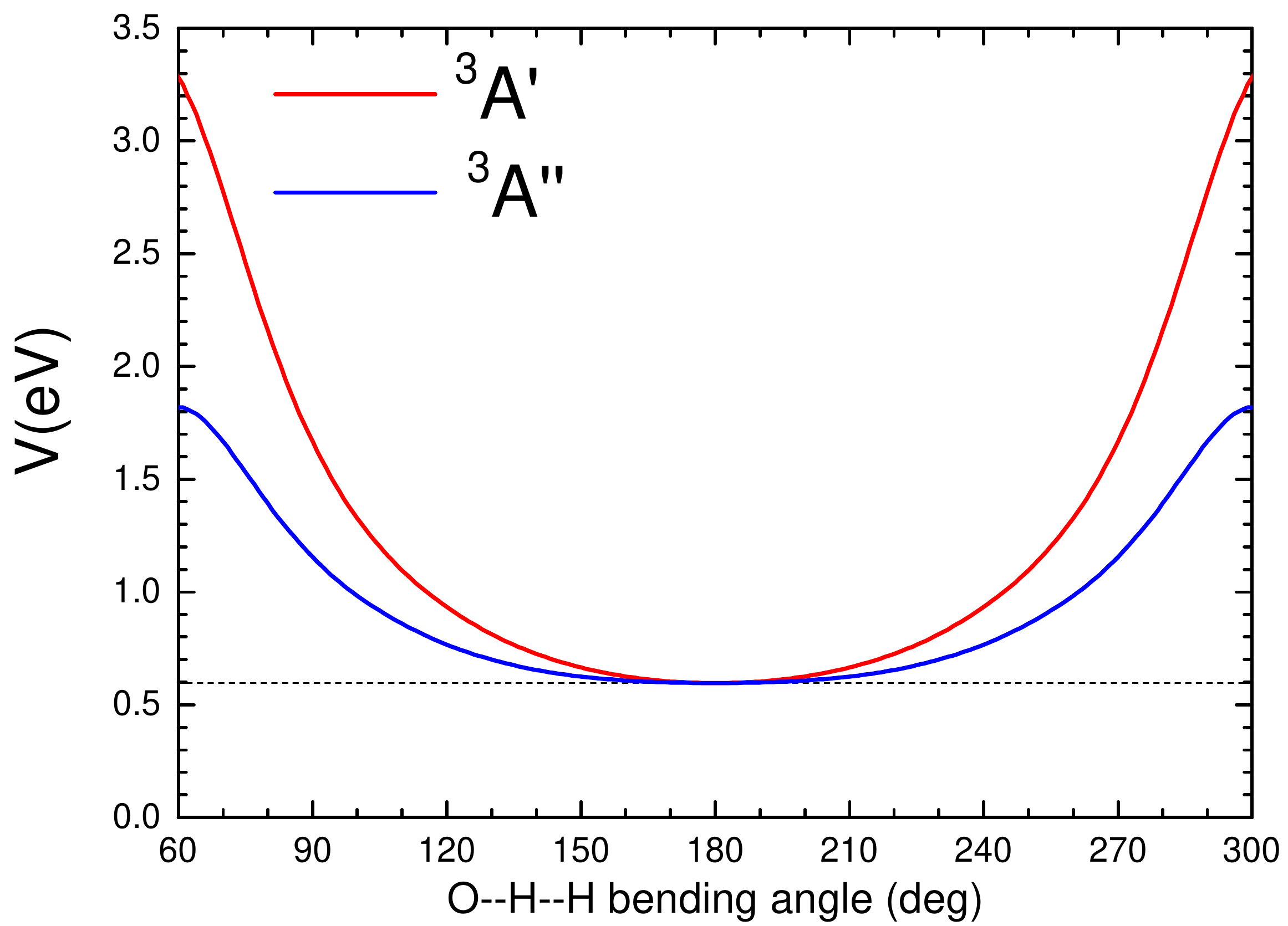}
\vspace*{0.5cm} \caption{Potential energy curves
of the $A'$ and $A''$ states as a function of the bending $\widehat{\rm HHO}$
angle. Each point in the figure correspond to the value of the
barrier height for the configuration at a given bending angle.}
\label{bending}
\end{figure}
%-----------------------------------------------------------------------

As the system draws away from the linear configuration, the degeneracy is
broken and hence the overall reactivity as well as other dynamical magnitudes
are expected to differ. Figure~\ref{bending} depicts the bending potential (the
height of the barrier as a function of the $\widehat{\rm OHH}$ angle) for the
pair of PESs. As can be seen, the $A'$ bending potential is considerably
steeper than that of the $A''$ PES, leading to a narrower cone of acceptance
and, as it will be shown, to a smaller reactivity. As a consequence of the
steeper bending potential on the $A'$ PES, the bending frequency on the $A'$ is
about a 70\% larger than on the $A''$ whereas, as expected, the transition
state symmetric stretch frequency is the same on both PESs, as shown in
Table~\ref{Saddle}. Therefore, the vibrational adiabatic potential (including
the ZPE) is 40 meV higher on the $A'$ PES, and this may have some
repercussion on the reaction threshold on this PES as compared to that on the
$A''$.

\subsection{Dynamical Results}
\label{Resdyn}

\begin{figure}
\includegraphics[scale=0.35,angle=0]{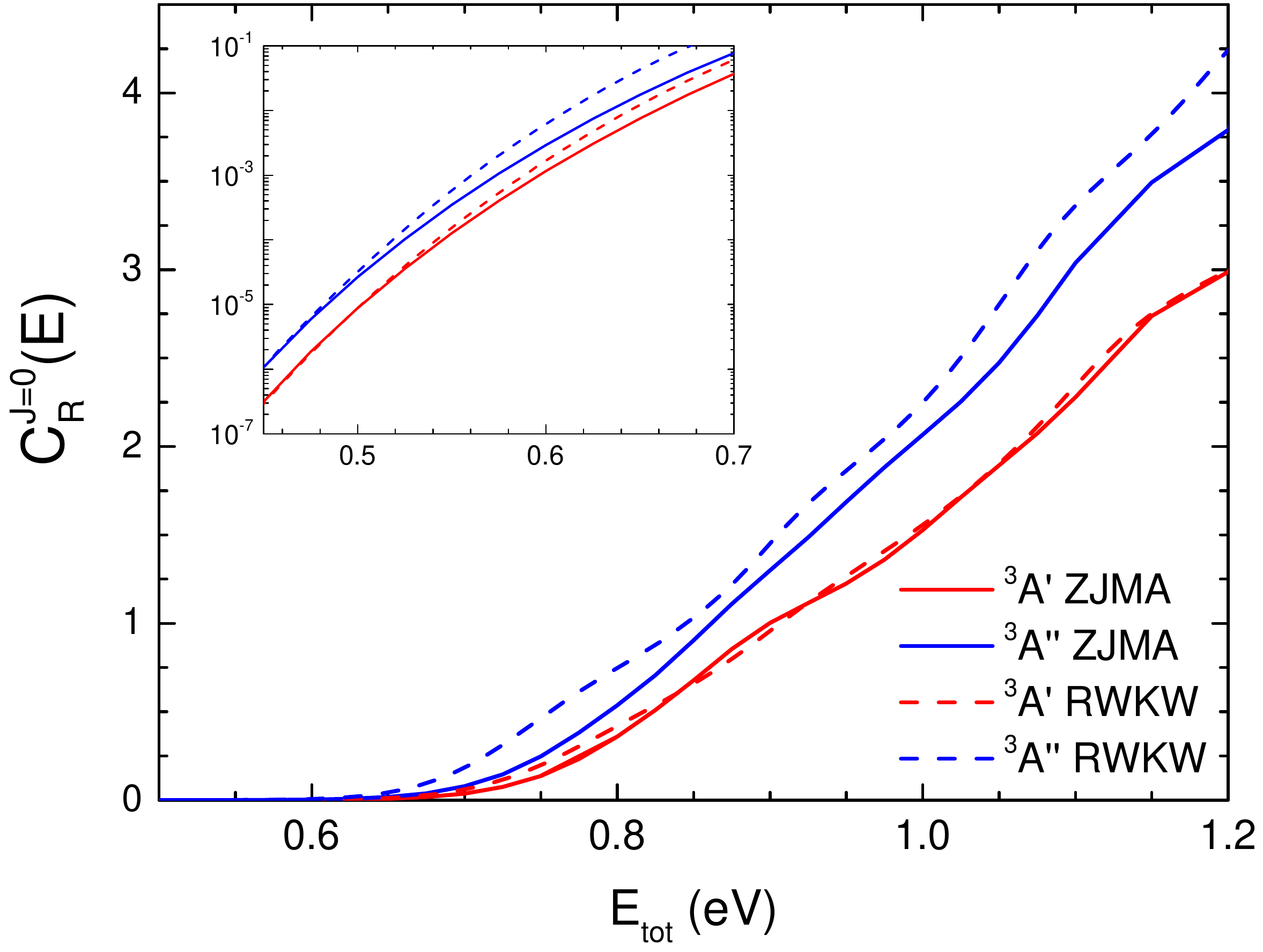}
\caption{QM Cumulative Reaction Probabilities
calculated for the $^3A'$ (red lines) and $^3A''$ (blue lines) states on the current, ZJMA PESs
(solid lines) and the RWKW PESs (dashed lines) for total angular momentum $J$=0.}
\label{crj0}
\end{figure}

The cumulative reaction probability\cite{M:JCP75,M:JCP76,M:ACR76,M:ACR93} (CRP), first
employed by Bill Miller in the context of a quantum version of the transition state
theory, can be considered as a measure of the number of reactant states that can proceed
to products for a given energy.  The CRP, which for its convenience is widely used, is
one of the most general, and yet accurate, dynamical quantities as it includes the
cumulative information for state-to-state processes from every reactant to every product
state. Furthermore, it is directly related to the reaction rate coefficient by simply
performing the average over the Boltzmann distribution of total energies
\cite{DFTCS:JACS91,DFTCS:F91,DFST:JPC92,JAEHS:JCP09,JLMLA:JCP12,AHS:JCP08,AHMS:PCCP07}.
Hence, it is an suitable dynamical quantity for comparing the overall reactivity of the
two competing $^3A'$ and $^3A''$ PESs involved. It can be used to compare the present
with other previous sets of PESs for the same reaction. The QM CRPs for total angular
momentum $J$=0, $C_R^{J=0}(E)$, calculated using the present PES (hereinafter ZJMA PES)
and the RWKW PESs by Rogers {\em et al.} are shown in Fig.~\ref{crj0}. As expected, based
on the evolution of the barrier height with the bending angle, the RWKW and ZJMA $A''$
PESs are more reactive than their $A'$ counterparts over the whole range of total
energies.

As can be seen, regardless of the total energy considered, the RWKW $A''$ PES
is systematically more reactive than the ZJMA $A''$ PES, while the CRPs for the
corresponding $A'$ PESs are fairly similar for total energies above 0.5 eV. At
lower energies the reaction probability is basically the same on the two sets
of PESs.  The lower reaction yield on the ZJMA PESs can be traced back to the
respective barrier heights on both sets of PESs; 24 meV for the $A'$ and 32 meV
for the $A''$ (see Fig.~\ref{meplinear}). It should also be stressed that, as
expected for reactions characterized by high early barriers, a step-like
pattern, associated to the opening of reactant's or products channels, does not
show up in any of the CRPs (at the energy necessary to surmount the barrier
there are already many open reactant's states). Furthermore, the lack of a
resonance pattern suggests that neither the entrance nor the exit Van der Waals
wells support bound states.

In what follows we will show only dynamical results calculating using the ZJMA
PESs.  As already mentioned, the large barrier and unfavorable kinematics for
OH detection has impaired experimental measurements for the title reaction.
Actually, most of the measurements have been carried out for the collisions
between oxygen and deuterium due to its somewhat  more favorable kinematics.
Relative excitation function was measured for the O($^3$P) + H$_2$ reaction
using a molecular beam apparatus and a hyperthermal O-atom beam
\cite{GMMTS:JCP03} and detection using mass spectrometry.
%--------------------------------------------------------------------------
\begin{figure}
\includegraphics[scale=0.45,angle=0]{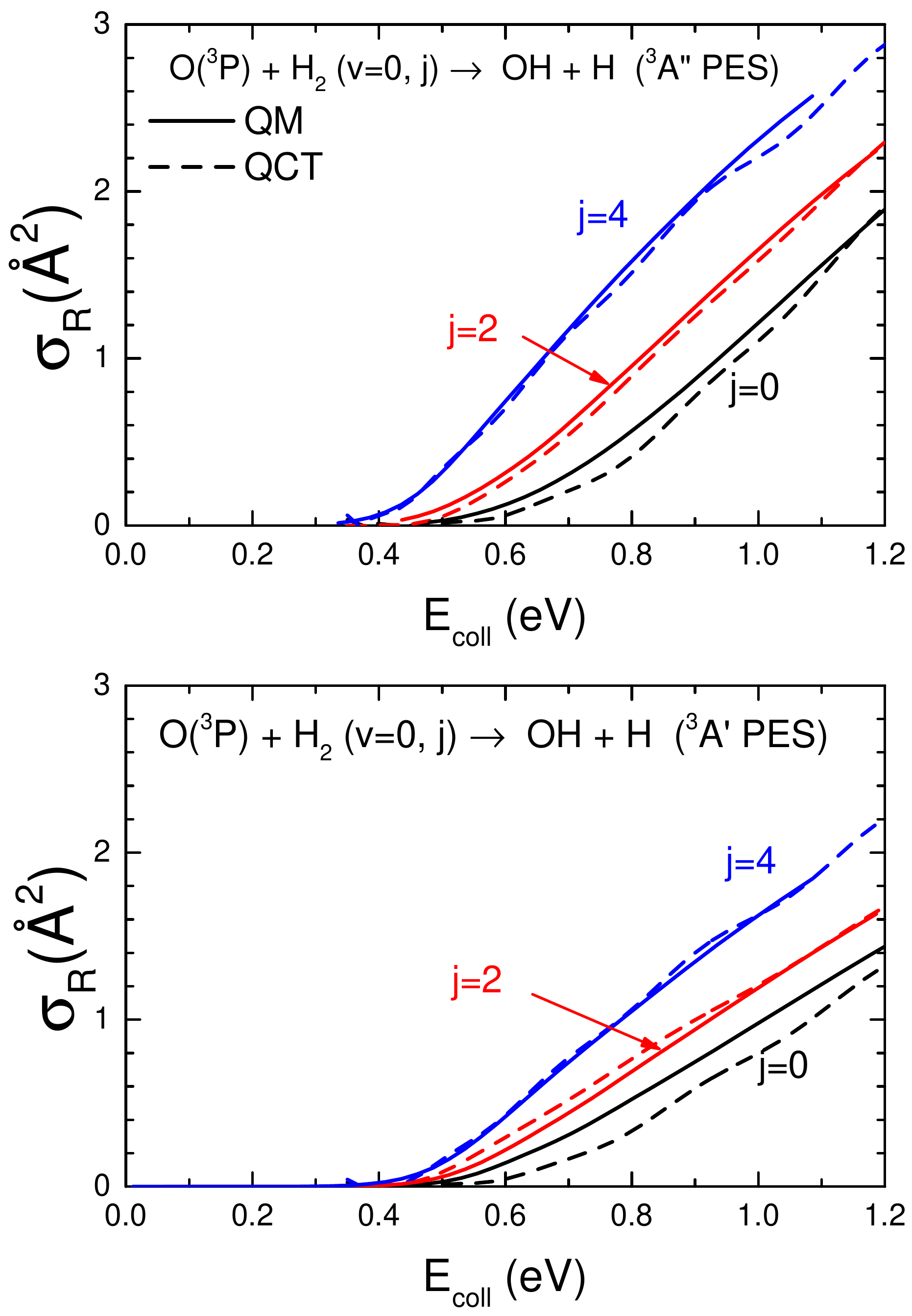}
\caption{ Comparison between the QCT excitation function, cross section as a
function of the collision energy, for the ($^3$P) + H$_2$ ($v$=0,$j$) reactions
on the $A''$ PES (top panel) and on the $A'$ PES (bottom panel). Results for
$j$=1, and 3 (not shown) follow the same trend than their even $j$
counterparts.}
\label{sigmarot}
\end{figure}
%--------------------------------------------------------------------------

To simulate the experimental excitation function, we have calculated them for different initial rotational
states ($j$=0-4) using fully converged QM hyperspherical and QCT methods. The results for
$j$=0, 2 and 4, displayed in Fig.~\ref{sigmarot}, show that regardless of the
collision energy considered, H$_2$ rotation always promotes the reactivity on the two
PESs. The agreement between QM and QCT results is fairly good, especially for $j$=2 and
4, although for $j$=0 the differences are appreciable, the QCT predicting lower cross
sections especially on the $A'$ PES. As expected from the previous studies, the cross
sections are smaller on the $A'$ than on the $A''$ PES for all initial rotational states.

%--------------------------------------------------------------------------
\begin{figure}
\includegraphics[scale=0.3,angle=0]{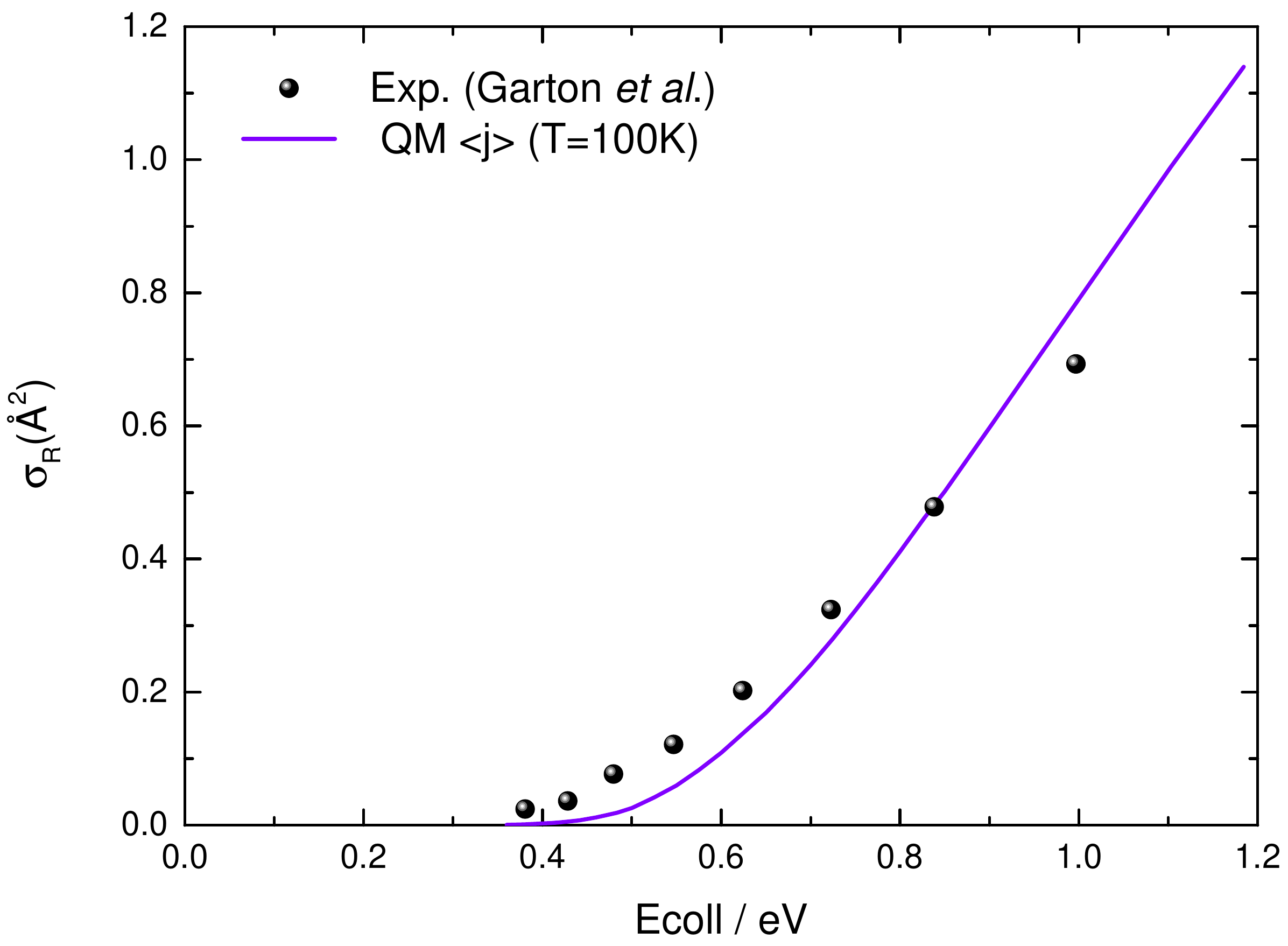}
\caption{ Comparison between the QM reactive cross sections calculated on the
current PESs and those measured by Garton \textit{et al.} \cite{GMMTS:JCP03}.
See text for more details.}
\label{compsigma}
\end{figure}
%
%------------------------------------------------------------------------------------
\begin{figure}
\includegraphics[scale=0.4,angle=0]{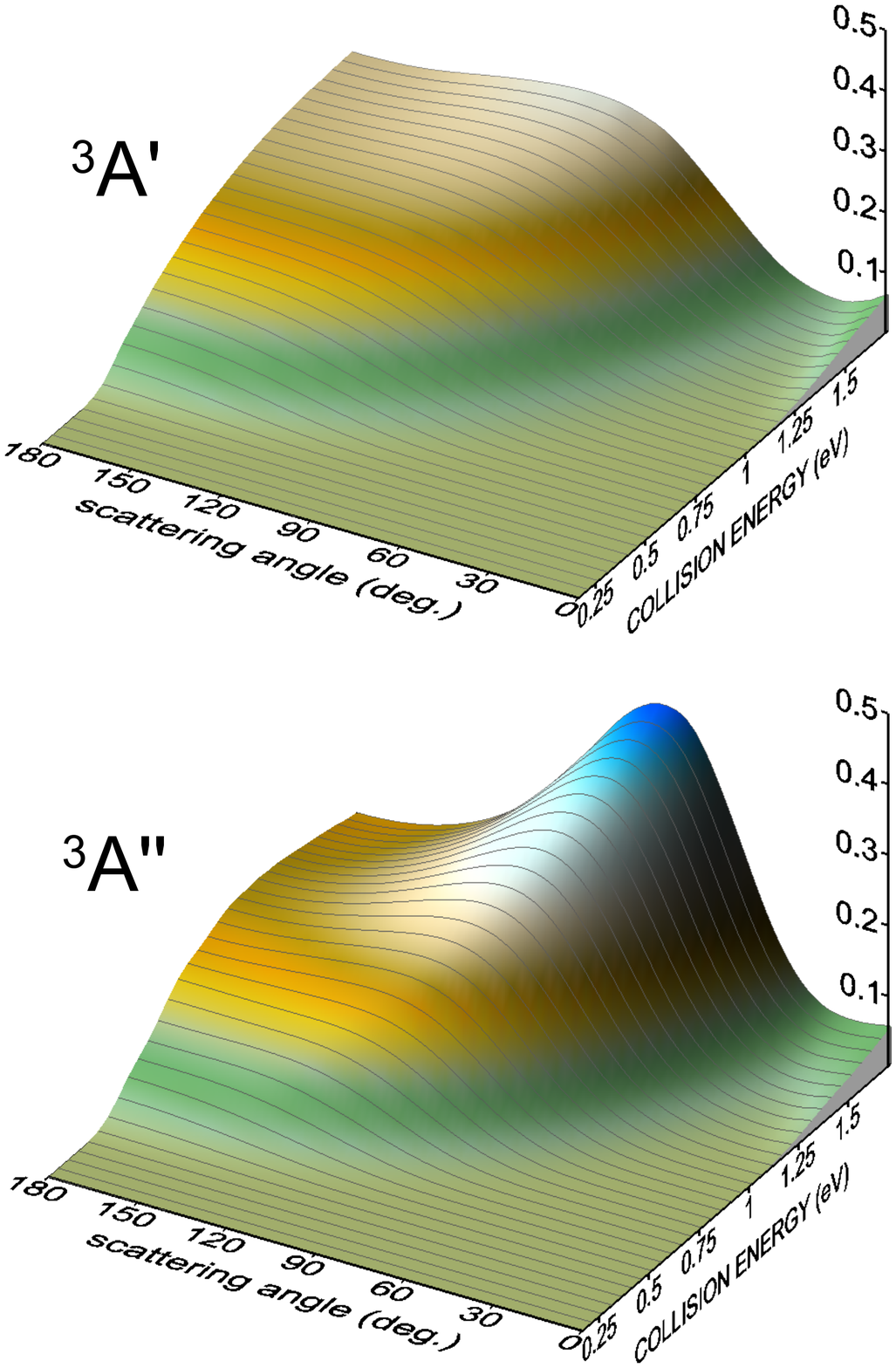}
\caption{  Reactive DCS as a function of the energy for the O($^3$P) + H$_2$
($v$=0,$j$=1) collisions on the $A''$ PES (top panel) and on the $A'$ PES
(bottom panel).}
\label{dcse}
\end{figure}
%--------------------------------------------------------------------------------
To account for the experimental conditions, QM excitation functions calculated on the
$A'$ and $A''$ PES were first averaged according to their statistical factors,
 \begin{equation}
   \sigma_R(j)=\frac{1}{3}\left(\sigma_R^{A'}(j) + \sigma_R^{A''}(j)  \right)
 \end{equation}
and then the excitation functions calculated for the different initial
rotational states were averaged according to a 100 K thermal distributions. The
results, displayed in Fig.~\ref{compsigma}, are compared to the experimental
excitation function.\cite{GMMTS:JCP03} Since no absolute values were obtained
in the cross beam experiments, their absolute values were obtained by scaling
the experimental measurements to the QM results. As can be seen, there is a
fairly good agreement except at energies close to the threshold, where the
present results on the ZMJA results underestimate the reactivity obtained in
the experiments. The difference can be in part attributed to contamination with
O($^1D$) in the beam. In Ref.~\citenum{GMMTS:JCP03} the authors state that the
population of O($^1D$) in the beam should be less than a 1\%, but reaction
between O($^1D$) and H$_2$ has no barrier and even such a small population of
O($^1D$) could alter the results at very low energies where the reaction yield
of O($^3P$) with H$_2$ is practically negligible.
%-------------------------------------------------------------------------
\begin{figure*}[t!]
\hspace*{-1.0cm}
\includegraphics[scale=0.6,angle=0]{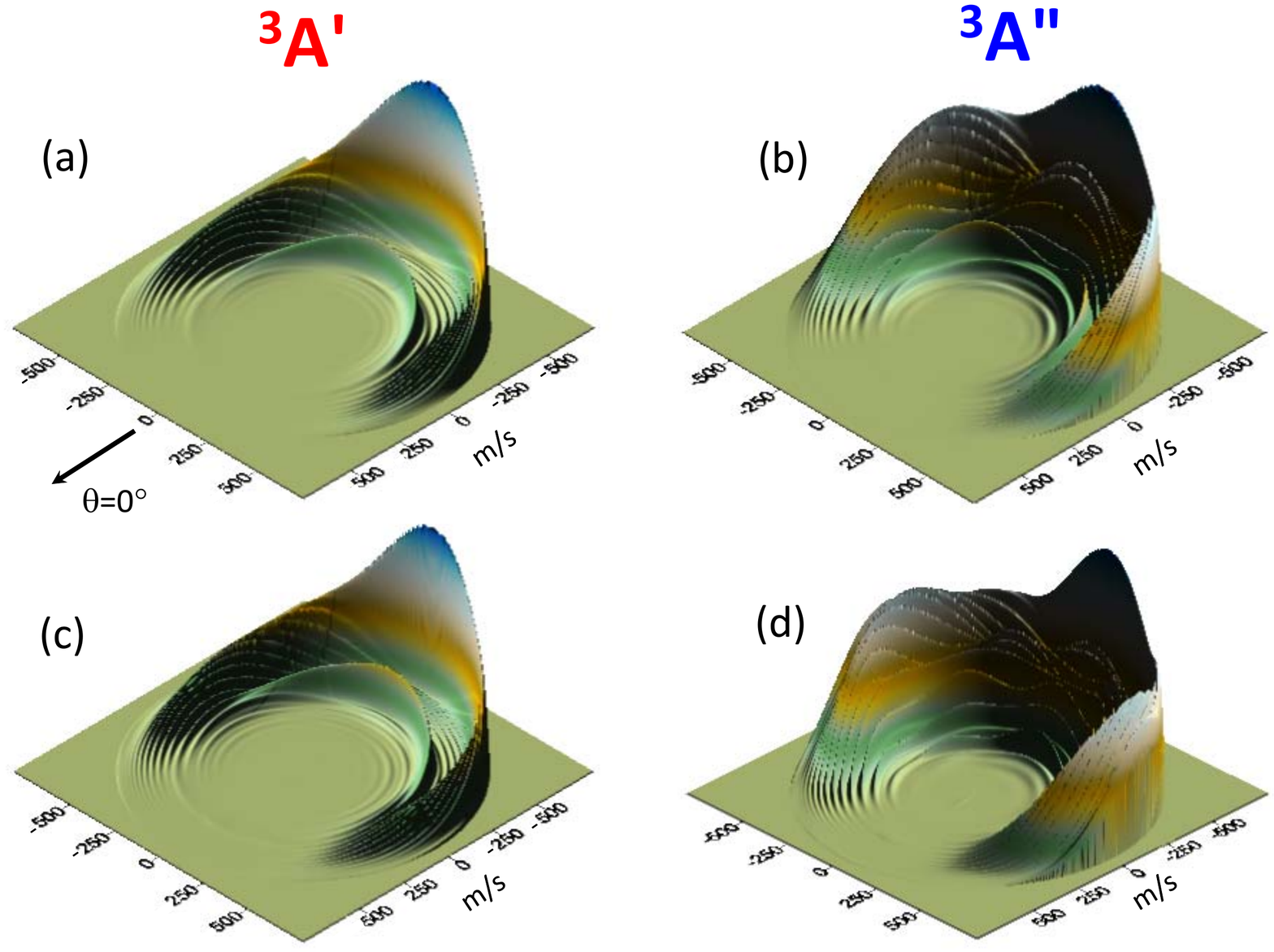}
\caption{Scattering angle-OH recoil velocity (ms$^{-1}$) triple differential cross
section (angle-velocity polar maps) at 0.83 eV (panels (a) and (b)) and 0.98 eV (panels
(c) and (d)) collision energies. Left and right panels depict the respective polar maps
on the $A'$ and $A''$ PES.}
\label{Fig10}
\end{figure*}
%----------------------------------------------------------------------------

To get a more detailed insight of the comparative reaction dynamics on both
PESs, we calculated the evolution  of the QM DCS with the collision energy. The
reaction with H$_2$($j$=1) was considered since it is the state that
contributes most to the thermal rate coefficient. These results, displayed in
Fig.~\ref{dcse},  show the different behavior for the non-collinear geometries
on the $A'$ and $A''$ PES causes not only changes in the absolute value, but
also the shape of the DCS. At low energies, the DCS on the $A'$ is confined in
the backward hemisphere showing a peak at 180$^\circ$. With increasing
collision energies, the peak transforms into a plateau that, at $E_{\rm
col}$=1.1 eV extends to 130$^\circ$. At $E_{\rm col}>$1.35 eV the DCS displays
a peak in the forward region and the maximum moves to 100$^\circ$. The DCS
calculated on the $A''$ covers a wider range of scattering angles. The most
salient feature is the shift towards sideways scattering  at energies above 1
eV, appearing as a prominent peak at $\theta$=90$^{\circ}$ beyond that value.
Although the ICS is consistently larger on the $A''$ PES, the DCS on the $A'$
PES is bigger at backward scattering angles and high collision energies.

The scattering angle-recoil velocity differential cross sections provide a
global view of the distribution of internal states as a function of the
scattering angle. Figure~\ref{Fig10} depicts  QM polar maps on the $A'$ (panels
a and c) and $A''$ (panels b and d) PESs for the O($^3P$)+H$_2(v=0, j=0$) at
0.83 eV (top panels) and 0.98 eV (bottom panels) collision energies. The main
difference between the results on both PESs is the presence of  evident
sideways scattering of the $A''$ PES that shifts towards higher angles with
increasing energy. By contrast, scattering on the $A'$ is confined to backward
angles at the two collision energies shown in Fig.~\ref{Fig10}. Moreover, the
rotational distribution is clearly colder on the $A'$ PES. Not only the maximum
of the distribution appears at lower $j'$ values, and hence at larger recoil
velocities, but also the rotational distribution is restricted to lower
rotational states. As a result of this, the ring corresponding to OH in $v'$=1
is clearly discernable on the $A'$ polar maps, clearly separated from the
rotational state rings of $v'$=0. The situation is less clear in the $A''$
polar maps, as a result of the partial overlap of the highest $v'=0, j'$ states
and the lowest $j'$ belonging to the $v'$=1 manifold.

\section{Conclusions}
\label{concl}

In this article we have presented a new set of global, analytically fitted PESs
for the triple state   O($^3$P) + H$_2$ reaction. Two adiabatic PESs of
symmetries $A'$ and $A''$ correlating with reactants and products, have been
calculated using MRCI and a large basis set. About 5000 ab initio points have
been calculated for each of the PES and fitted using the using the GFIT3C
procedure. The resulting fits reproduce the {\it ab initio} potential with
great accuracy. In particular, the degeneracy of the transition state region
for the O-H-H collinear configuration is reproduced almost exactly. This makes
the new set of potential especially suitable for studies aimed at comparing the
dynamics on the  $A'$ and the $A''$ PESs. In addition, the long range potential
in the entrance channel is also considered, so that these new PESs could also
be used to study collisional excitation of H$_2$ at low temperature.

Quantum mechanical, fully converged, calculations have been performed on both
PESs.  In good agreement with previous studies, the $A''$ PES is found
considerably more reactive, a behavior that can be traced back to its broader
cone of acceptance. Calculations of the excitation functions (collision energy
dependence of the reaction cross section) have been carried out for all the
rotational states from $j$=0 to $j$=4. A satisfactory agreement has been found
in the comparison of the experimental and theoretically simulated excitation
functions using the calculations on  both PESs weighted over the initial
rotational distribution. Differential cross sections (DCSs) as a function of
the collision energy are also shown in  the present work. The distinct
topography of the $A'$ and $A''$ PESs is reflected in the DCS, which displays a
prominent sideways peak on the $A''$ at energies above 1.0 eV  that is absent
on the $A'$ PES. In addition, scattering angle-OH recoil velocity polar maps
have been calculated at two collision energies to illustrate in a graphical
manner the considerable difference in the dynamics of the reaction on each of
the PES.

\section{Acknowledgment}

The authors thank Prof. Enrique Verdasco for his help with the calculations.
Funding by the Spanish Ministry of Science and Innovation (grant
MINECO/FEDER-CTQ2015-65033-P, and PGC2018-09644-B-100) is also acknowledged.
P.G.J. acknowledges funding by Fundaci\'on Salamanca City of Culture and
Knowledge (programme for attracting scientific talent to Salamanca).

%% ****** Start of file sortemplate.tex ****** %
%%
%%   This file is part of the files in the distribution of AIP substyles for REVTeX4.
%%   Version 4.1 of 9 October 2009.
%%
%
% This is a template for producing documents for use with
% the REVTEX 4.1 document class and the AIP substyles.
%
% Copy this file to another name and then work on that file.
% That way, you always have this original template file to use.

%\preprint{}
\draft % marks overfull lines with a black rule on the right

\newpage

\end{document}